\pdfoutput=1
\documentclass[aps,prb,reprint,superscriptaddress]{revtex4-1}

\usepackage{textcomp}
\usepackage{array}
\usepackage{graphicx}
\usepackage{float}
\usepackage{amsmath}
\usepackage{amssymb}
\usepackage{epstopdf}
\usepackage{bm}
\usepackage{array}
\newcolumntype{P}[1]{>{\centering\arraybackslash}p{#1}}

\begin{document}

\title{Stable quantum dots in an InSb two-dimensional electron gas}

\author{Ivan~Kulesh}
\affiliation{QuTech and Kavli Institute of Nanoscience, Delft University of Technology, 2600 GA Delft, The Netherlands}

\author{Chung~Ting~Ke}
\affiliation{QuTech and Kavli Institute of Nanoscience, Delft University of Technology, 2600 GA Delft, The Netherlands}

\author{Candice~Thomas}
\affiliation{Department of Physics and Astronomy, Purdue University, West Lafayette, Indiana 47907, USA}
\affiliation{Birck Nanotechnology Center, Purdue University, West Lafayette, Indiana 47907, USA}

\author{Saurabh~Karwal}
\affiliation{QuTech and Netherlands Organization for Applied Scientific Research (TNO), 2628 CK Delft, The Netherlands}

\author{Christian~M.~Moehle}
\affiliation{QuTech and Kavli Institute of Nanoscience, Delft University of Technology, 2600 GA Delft, The Netherlands}

\author{Sara~Metti}
\affiliation{Birck Nanotechnology Center, Purdue University, West Lafayette, Indiana 47907, USA}
\affiliation{School of Electrical and Computer Engineering, Purdue University, West Lafayette, Indiana 47907, USA}

\author{Ray~Kallaher}
\affiliation{Birck Nanotechnology Center, Purdue University, West Lafayette, Indiana 47907, USA}
\affiliation{Microsoft Quantum Purdue, Purdue University, West Lafayette, Indiana 47907, USA}

\author{Geoffrey~C.~Gardner}
\affiliation{Birck Nanotechnology Center, Purdue University, West Lafayette, Indiana 47907, USA}
\affiliation{Microsoft Quantum Purdue, Purdue University, West Lafayette, Indiana 47907, USA}

\author{Michael~J.~Manfra}
\affiliation{Department of Physics and Astronomy, Purdue University, West Lafayette, Indiana 47907, USA}
\affiliation{Birck Nanotechnology Center, Purdue University, West Lafayette, Indiana 47907, USA}
\affiliation{School of Electrical and Computer Engineering, Purdue University, West Lafayette, Indiana 47907, USA}
\affiliation{Microsoft Quantum Purdue, Purdue University, West Lafayette, Indiana 47907, USA}
\affiliation{School of Materials Engineering, Purdue University, West Lafayette, Indiana 47907, USA}

\author{Srijit~Goswami}\email{S.Goswami@tudelft.nl}
\affiliation{QuTech and Kavli Institute of Nanoscience, Delft University of Technology, 2600 GA Delft, The Netherlands}

\begin{abstract}

Indium antimonide (InSb) two-dimensional electron gases (2DEGs) have a unique combination of material properties: high electron mobility, strong spin-orbit interaction, large Land\'{e} g-factor, and small effective mass. This makes them an attractive platform to explore a variety of mesoscopic phenomena ranging from spintronics to topological superconductivity. However, there exist limited studies of quantum confined systems in these 2DEGs, often attributed to charge instabilities and gate drifts. We overcome this by removing the  $\delta$-doping layer from the heterostructure, and induce carriers electrostatically. This allows us to perform the first detailed study of stable gate-defined quantum dots in InSb 2DEGs. We demonstrate two distinct strategies for carrier confinement and study the charge stability of the dots. The small effective mass results in a relatively large single particle spacing, allowing for the observation of an even-odd variation in the addition energy. By tracking the Coulomb oscillations in a parallel magnetic field we determine the ground state spin configuration and show that the large g-factor ($\sim$30) results in a singlet-triplet transition at magnetic fields as low as 0.3~T.

\end{abstract}

\maketitle

Mesoscopic devices can be made in a two-dimensional electron gas (2DEG) using electrical gates to confine charge carriers, thus reducing the degrees of freedom. The extreme case is the zero-dimensional system, a quantum dot (QD). QDs have been used to explore a wide range of quantum phenomena\cite{kouwenhoven1997electron,Reimann2002,Hanson2007}, and are emerging as a platform for quantum computing\cite{Loss1998,Petta2005,Veldhorst2015} and quantum simulations\cite{Byrnes2008,Manousakis2002,Hensgens2017}. QDs in materials with strong spin-orbit interaction have important applications in the field of topological superconductivity. They can be used to realize Majorana zero modes in quantum dot chains\cite{sau2012realizing}, and are essential elements for the readout and manipulation of topological qubits~\cite{Karzig2016,Plugge2017}.

InSb is a promising material in this regard, with a high carrier mobility, large g-factor and strong spin-orbit interaction\cite{Lei2019, Qu2016, Kallaher2010, nedniyom2009giant}. QDs have been extensively studied in InSb nanowires\cite{Nilsson2009,Nadj-Perge2012a,VanWeperen2015} and, more recently, in InSb nanoflakes\cite{Xue2019}. However, despite the clear benefits of scalability offered by InSb 2DEGs, experimental reports of confined systems in these quantum wells are scarce~\cite{Orr2007,Qu2016,Masuda2018}. Thus far, most studies of InSb 2DEGs have been limited to heterostructures where carriers are generated in the quantum well via remote $\delta$-doping layers. Such doped 2DEGs are known to suffer from charge instability and gate drifts, which are particularly detrimental to the study of nanostructures.

Here, we show that removal of doping layers enables the realization of highly stable gate-defined quantum dots in InSb 2DEGs. We study two different QD designs on deep and shallow undoped quantum wells. In all measured devices, the charge stability diagrams show well-defined Coulomb diamonds and excited states. We find that the small effective mass of InSb results in a large separation between single-particle levels, enabling the observation of an even-odd periodicity in the Coulomb peak spacing. Studying the evolution of these peaks in magnetic field allows us to extract the g-factor and determine the parity of ground states.

The QDs are fabricated on InSb quantum wells grown by molecular beam epitaxy on GaAs (100) substrates. Electronic confinement in the quantum well is obtained with the growth of AlInSb barriers on either side, see Fig.~\ref{fig1}(a). Further details on the growth, layer stack and characterization of wafers are provided in the supplementary information (SI). The key difference here, as compared to previous studies\cite{Orr2007,Yi2015,Qu2016,Masuda2018,Lei2019}, is the absence of a $\delta$-doping layer which is typically inserted above the quantum well to generate carriers. Instead, we populate the well by applying positive voltages to an accumulation gate, an approach widely used in other semiconductor 2DEG materials\cite{lu2011enhancement, borselli2011pauli, hendrickx2018gate}.

Figure~\ref{fig1}(b), (d) show schematics of QDs with either a single or double layer of gates. The fabrication flow is similar for both types of designs (for more details see SI). We first isolate lithographically defined mesas using wet chemical etching. Ohmic contacts are made to the buried quantum well after removal of the native oxide via sulphur passivation\cite{Gong1997,Zhang2017}. The entire structure is then covered by 40~nm of aluminum oxide with atomic layer deposition (ALD), followed by the deposition of top gates. Figure~\ref{fig1}(c), (e) show scanning electron micrographs at this stage for the single and double layer devices respectively. While the fabrication of single-layer QDs ends here, the double-layer devices require an additional layer of aluminum oxide, followed by a global accumulation gate (A).

\begin{figure}[t!]
	\includegraphics[scale=1]{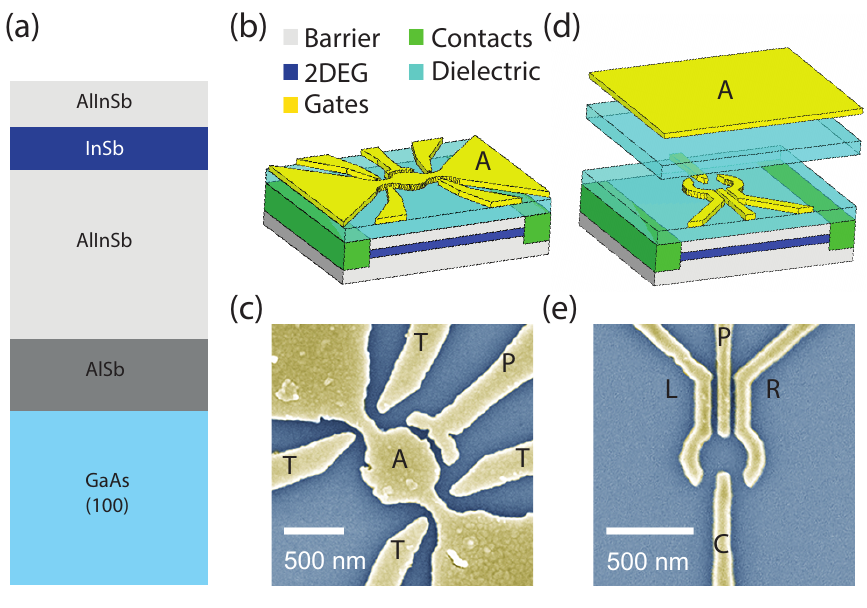}
	\caption{(a) Layer stack of the InSb/AlInSb heterostructure. (b) Sketch of the single-layer and (d) the double-layer types of design. Positive voltage is applied to gates A populating charge carriers, other gates are energised with negative voltages and control the system as explained in the text. (c) False-colored scanning electron micrograph of the single-layer dot and (e) the double-layer dot prior to the second gate layer deposition.}
	\label{fig1}
\end{figure}

The single-layer QD has a local accumulation gate (A), which when energized with a positive voltage, creates a populated 2DEG region following the shape of the gate. The tunnel gates (T) are then used to tune the barriers, and the plunger gate (P) controls the chemical potential of the dot. For double-layer QDs carriers are induced via the global accumulation gate. The quantum dot confinement potential and tunnel barriers are controlled via the fine gates labeled left (L), right (R), and central (C), and the plunger (P) again tunes the chemical potential. While the first design is simpler and offers the flexibility to integrate with other mesoscopic systems, the latter provides better control on the dot size and occupation with the possibility of depleting it to the few electron regime\cite{ciorga2000addition}. Here we report results on three quantum dots in undoped wafers (S1, S2, S3) and one on a doped wafer (S4) for comparison. The relevant design type and top-barrier thickness are summarized in Table~\ref{table1}. All measurements are performed in a refrigerator with a base temperature of 300~mK. We use standard lock-in techniques with a AC excitation of 10$\,-\,$20~$\mu$V, and measure the differential conductance $G$.

\begin{figure}[!b]
	\includegraphics[scale=1]{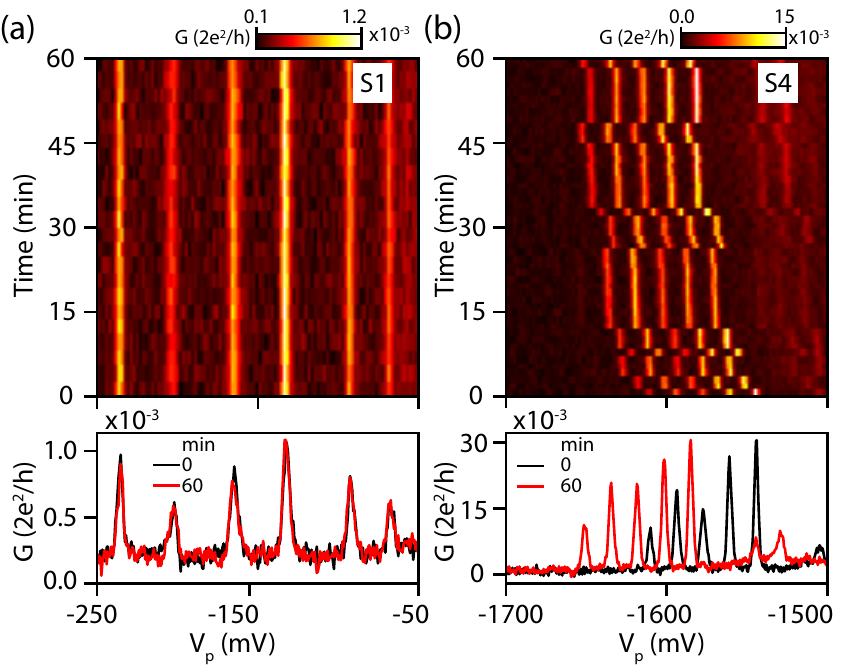}
	\caption{(a) Top: stability of coulomb oscillations as a function of time for undoped wafer, sample S1. Bottom: the line cuts for t$\,=\,$0 (black) and t$\,=\,$ 60 (red) minutes. Gate settings: V$_\text{L} = -202$~mV, V$_\text{R} = -252$~mV, V$_\text{C} = -145$~mV, V$_\text{A} = +500$~mV. (b) Same measurement for the doped wafer, sample S4. Gate settings: V$_\text{L} = -1600$~mV, V$_\text{R} = -1500$~mV, V$_\text{C} = -1630$~mV.}
	\label{fig2}
\end{figure}

We first compare the stability of QDs fabricated on undoped (S1) and doped (S4) quantum wells. Both devices are fabricated simultaneously with an identical gate geometry. Note that S4 does not require the global accumulation gate, but the design of fine gates is similar to the one in Fig.~\ref{fig1}(e). The QDs are tuned to the Coulomb blockade regime, and we monitor the Coulomb oscillations as a function of time. Figure~\ref{fig2}(a) shows that S1 is extremely stable in time. On the other hand, S4 shows a drift and jumps in the peak position, associated with charge instabilities in the device (Fig.~\ref{fig2}(b)). We observe similar characteristics in a variety of doped wafers and here only present data from the most promising QD. In contrast, we show here that stable QDs can be reliably fabricated in undoped heterostructures, irrespective of the 2DEG depth or the specific QD design.

\begin{figure}[!b]
	\includegraphics[scale=1.0]{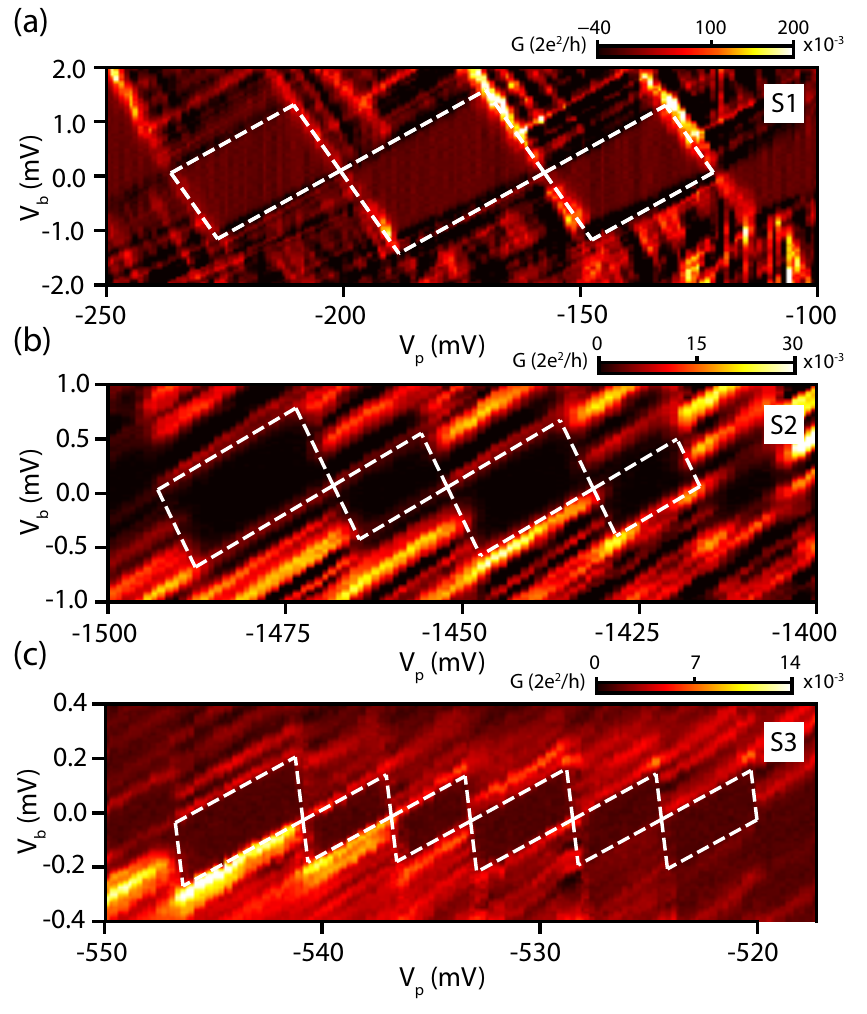}
	\caption{Charge stability diagrams for (a) S1, (b) S2, and (c) S3. Coulomb blockaded regions are highlighted with white dashed lines and are used to estimate the charging energies.}
	\label{fig3}
\end{figure}

In Fig.~\ref{fig3} we plot $G$ as a function of the bias voltage ($V_\text{b}$) and plunger gate voltage ($V_\text{P}$) for devices S1-S3. S1 is a double-layer QD in a deep (40~nm) 2DEG, and S2/S3 are double/single-layer QDs fabricated on a shallow (10~nm) 2DEG. The charge stability diagrams (typically acquired over several hours) are devoid of charge jumps and show well-defined Coulomb diamonds. Transport through excited states is visible in the form of conductance peaks running parallel to the diamond edges. We also observe regions with a negative differential conductance (particularly clear for S1), possibly arising from suppressed transport through specific excited states of the QD~\cite{weis1993competing}. From the size of the smallest diamond we estimate charging energies E$_\text{c}$ of 1.23~meV (S1), 0.5~meV (S2) and 0.16~meV (S3). These values are in good agreement with the designed geometries (see SI for estimates of the QD size). To fully characterize the dots we also perform detailed gate vs. gate measurements for different combinations of gates for each of the devices (shown in SI).
For double-layer dots, the left and right gates couple to the QD equally (C$_\text{L}$/C$_\text{R}\, \approx\,0.9$ for S1 and C$_\text{L}$/C$_\text{R}\, \approx\,1$ for S2) and more strongly than the plunger gate (C$_\text{L}$/C$_\text{P}\, \approx\,3.8$ for S1 and C$_\text{L}$/C$_\text{P}\, \approx\,2.6$ for S2). For the single-layer dot S3 the tunnel gate T, with only tips in the vicinity of the active region, has smaller coupling compared to the plunger gate, C$_\text{T}$/C$_\text{P}\, \approx\,0.6$. These coupling ratios agree with expectations from the respective QD designs, confirming the realization of well-defined quantum confinement.

\begin{table}[!t]
	\centering
	\begin{tabular}{ |P{1.2cm}||P{1.2cm}|P{2.4cm}|P{1.7cm}|}
		\hline
		device & doping & QW depth (nm) & gate layers \\
		\hline
		S1	& No & 40 & 2 \\
		S2	& No & 10 & 2 \\
		S3	& No & 10 & 1 \\
		S4	& Yes & 40 & 1 \\
		\hline
	\end{tabular}
	\caption{Overview of samples studied in this work.}
	\label{table1}
\end{table}

\begin{figure}[!t]
	\includegraphics[scale=1]{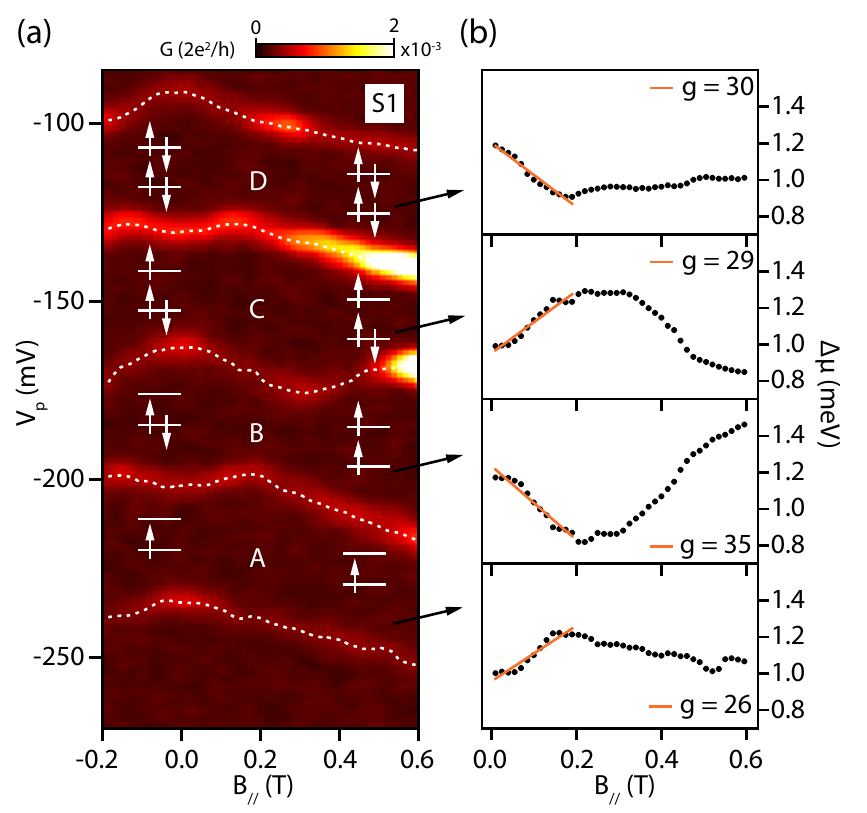}
	\caption{(a) Evolution of Coulomb blockade peaks for S1 as a function of in-plane magnetic field $B_{||}$. White dashed lines mark the peak positions and arrows represent the ground state spin configuration. (b) Extracted addition energy as a function of $B_{||}$. The g-factor is extracted from a linear fit (orange line) in the low field regime.}
	\label{fig4}
\end{figure}

The stability diagram in Fig.~\ref{fig3}(a) also reveals an even-odd variation in the diamond size for S1, consistent with the spin-dependent filling of the orbital levels \cite{Cobden2002,Nilsson2009,fasth2007direct}. The low effective mass in InSb 2DEGs\cite{Ke2019,Lei2019} allows us to observe this effect in relatively large QDs\cite{blanter1997fluctuations}. We confirm the alternating spin filling by studying the response of Coulomb peaks to a magnetic field applied in the plane of the 2DEG ($B_{||}$). In Fig.~\ref{fig4}(a), we show the magnetic field evolution of five Coulomb blockade peaks for S1 from $B_{||} =-0.2$~T to 0.6~T. The corresponding ground states (GSs) are labeled A-D, and peak positions (indicated by white dashed lines) 
are determined by fitting a Gaussian function. We note that the gate configuration here is different as compared to 
Fig.~\ref{fig3}(a). The corresponding diamonds are presented in the SI. At low fields between ${-0.2}$ to 0.2~T, successive Coulomb peaks move in opposite directions. This is consistent within the non-interacting picture where the consecutive filling of electrons is based on the Pauli exclusion principle. Starting with an empty level, the first electron fills with a spin up/down, then the second electron fills the same state with opposite spin. In this case, one expects two consecutive Coulomb blockade peaks to move apart with increasing Zeeman field. However, when the next electron enters the dot, it has to occupy a higher quantum level. Therefore, for two consecutive electrons belonging to different quantum levels, corresponding peaks move toward each other as the field increases.

Figure~\ref{fig4}(a) can also be represented in terms of the addition energy, defined as the difference in chemical potential ($\mu$) between successive ground state transitions, i.e., $\Delta \mu = \mu_{\text{N}+1} - \mu_{\text{N}}$. This value can be extracted directly from the Coulomb peak spacing by converting gate voltage to energy using the plunger lever arm (see SI). As shown in Fig.~\ref{fig4}(b), a linear region is observed with addition energy proportional to the Zeeman term, $\pm\,\text{g}\mu_\text{B}B$ where the sign depends on the parity of ground state. This allows us to extract the absolute value of the g-factor, which lies in the range 26 -- 35 for the four GSs analyzed here. At higher magnetic fields ($B_{||}\simeq 0.3$~T), states B and C display an overturned behavior, corresponding to a triplet ground state with total spin of 1, rather than a singlet state with total spin of zero. This singlet to triplet transition is expected when the Zeeman energy is comparable to the singlet-triplet gap at the zero field\cite{Hanson2007, fasth2007direct, Nilsson2009, Deon2010}. It is worth noting that the large g-factor of InSb 2DEGs allows for the observation of these ground state transitions at significantly lower magnetic fields than many other material systems~\cite{Potok2003,fasth2007direct,katsaros2010hybrid,Kurzmann2019}.

In conclusion, we demonstrate the successful realization of stable, controllable quantum dots in InSb 2DEGs. This stability allows us to fully characterize dots fabricated using two distinct designs. We show that the low effective mass leads to spin-dependent filling of the quantum levels for relatively large quantum dots. Furthermore, we extract a large Land\'{e} g-factor ($\sim 30$), which results in a singlet-triplet transition at low magnetic fields. Our studies show that InSb quantum wells are an excellent platform to study quantum confined systems, and particularly relevant for future applications in topological superconductivity.
\\

\textbf{Acknowledgments}
We thank Fokko de Vries and Klaus Ensslin for comments on the manuscript, and Jasper van Veen for helpful discussions.
The research at Delft was supported by the Dutch National Science Foundation (NWO), the Early Research Programme of the Netherlands Organisation for Applied Scientific Research (TNO) and a TKI grant of the Dutch Topsectoren Program. The work at Purdue was funded by Microsoft Quantum.
\\

\textbf{Data availability}
The data sets presented in the figures are available at: https://doi.org/10.4121/uuid:28a121af-1e08-429d-9d53-1cc53764e91e


\bibliography{InSbQD}

\end{document}


\title{Supplementary Information for: Stable quantum dots in an InSb two-dimensional electron gas}


\section*{\large{Supplementary Information}}
\vspace{10mm}

\section{Wafer description}

\begin{figure}[h!]
	\includegraphics[scale=1]{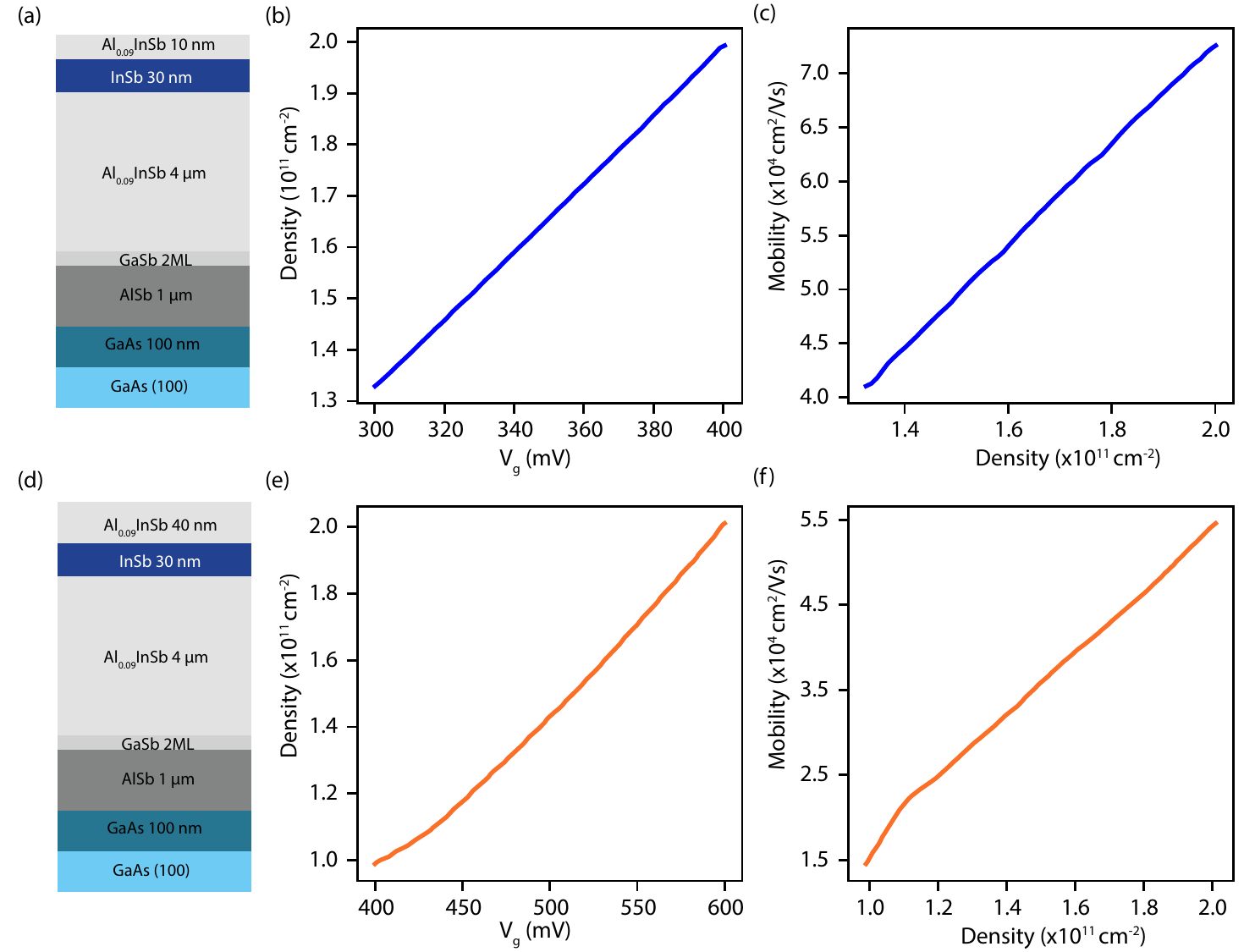}
	\caption{Characterization of the two undoped InSb quantum wells used in this study, using Hall bars. The top (bottom) row corresponds to a wafer with top barrier thickness of 10~nm (40~nm). Details of the growth process are similar to previous work \cite{Ke2019}, the only difference being the absence of the $\delta$-doping layer in these heterostructures.  (a,d) Schematics of the wafer stack. (b,e) Carrier density as a function of voltage on the accumulation gate. (c,f) Mobility as a function of density.}
	\label{fig1}
\end{figure}

\clearpage
\section{Accumulation curves}

\begin{figure}[h!]
	\includegraphics[scale=1]{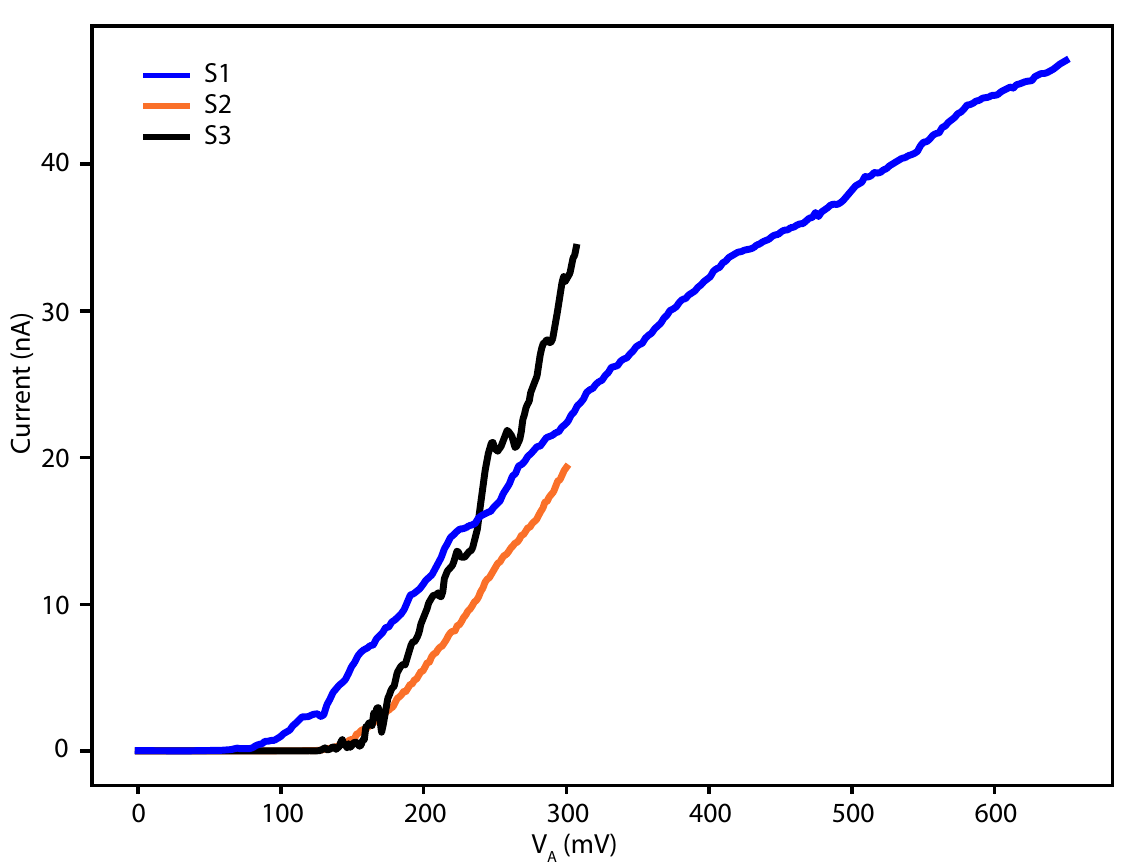}
	\caption{Current through the sample as a function of voltage applied to the accumulation gate, while all other gates are grounded. Applied bias voltage is 1~mV. The range is limited by the onset of gate leakage.}
	\label{fig1_2}
\end{figure}

\clearpage
\section{Fabrication details}
In this section the fabrication steps are presented, used for all quantum dots and Hall bar devices discussed.

\begin{itemize}

\item We define mesa regions with e-beam lithography, using PMMA A4 950 resist. Prior to exposure, the resist is baked for 60 min in a vacuum oven at $100\, ^\circ$C. Using resist as a mask, we etch away unprotected regions using a citric acid solution (560 ml $\text{H}_2\text{O}$, 5~ml 37\% $\text{H}_2\text{O}_2$, 7~ml 50\% $\text{H}_3\text{P}\text{O}_3$, 9.6~g citric acid). The etch depth is approximately 50~nm below the quantum well.

\item Contact leads are defined with PMMA A6 950 resist. Prior to ohmic contact deposition, the native oxide at the contact interface is removed by sulphur passivation in an aquatic solution of 0.5\% ammonium polysulfide at  60 $^\circ$C for 30 min \cite{Gong1997,Zhang2017}. The sample is transferred to the evaporator chamber in a water beaker to reduce exposure to the air. First, the sample is gently etched in He plasma to remove sulphur residues, followed by evaporation of 10~nm of Ti and 150~nm of Au.

\item We deposit $\approx$ 40 nm $\text{Al}_2\text{O}_3$ using ALD at $105^\circ$C. The dielectric layer isolates the top gates from mesa and ohmic contacts. Moreover, it improves the gating performance compared to Schottky barriers between the metal gates and InSb 2DEGs \cite{uddin2013gate}.

\item Top gates are evaporated in two steps to improve resolution. Fine structures are defined using PMMA A4 950 resist and 10 nm Ti / 50 nm Au are evaporated. Fine gates do not cross mesa borders. Coarse gates are defined using bilayer PMMA A6 495 / PMMA A3 950. Then we evaporate 10 nm Ti / 150 nm Au, allowing for the gates to climb the mesa edges.

\item When using the double layer design, ALD and coarse top gates steps are repeated.

\end{itemize}

\clearpage
\section{Gates cross-coupling characterization}

\begin{figure}[h!]
	\includegraphics[scale=1]{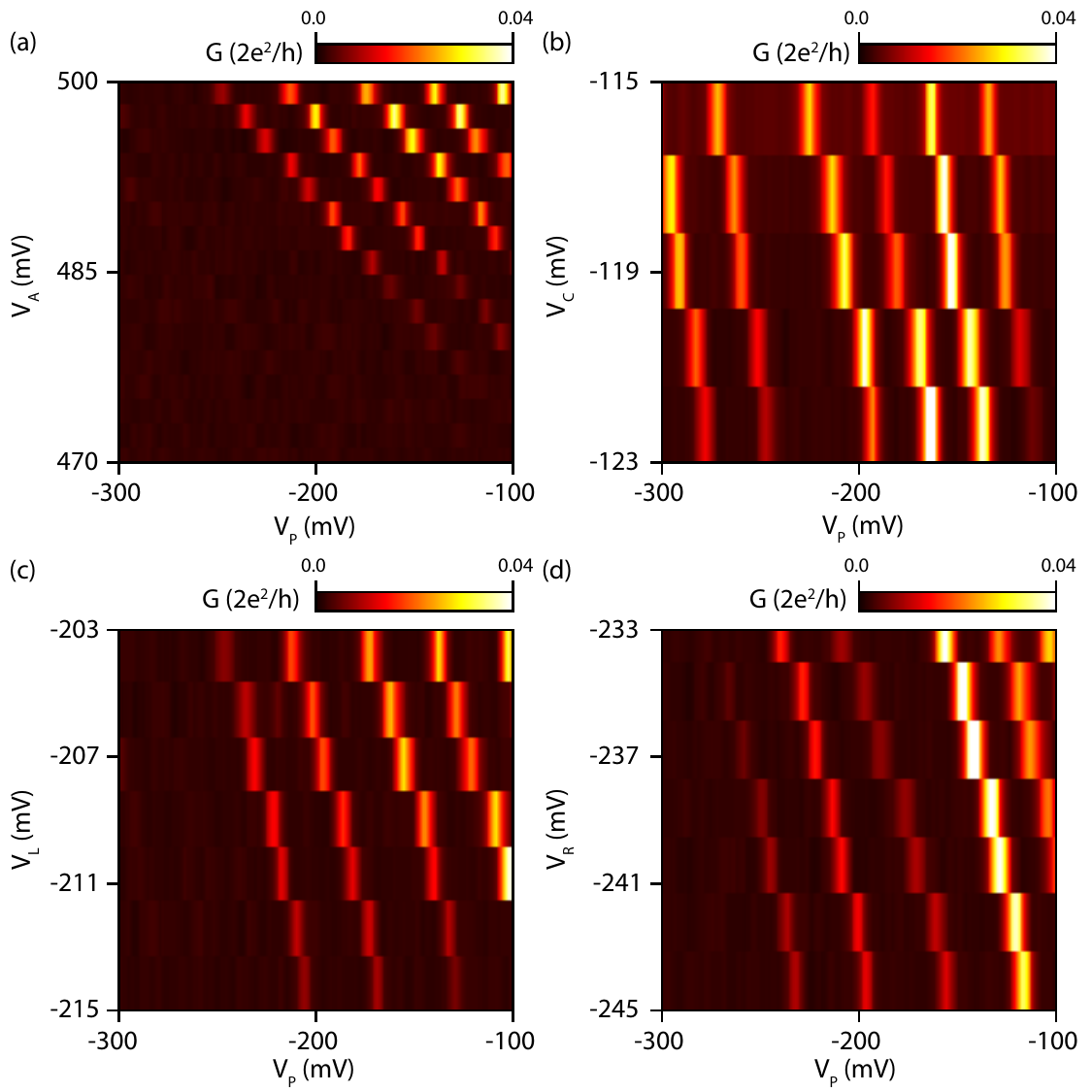}
	\caption{Conductance maps of the S1 double-layer dot at zero bias for different combinations of gates. Coulomb oscillations slopes are used to determine gates coupling ratios. (a) Accumulation gate vs. plunger gate: C$_\text{A}$/ C$_\text{P}\,=\,5.5$. (b) Central gate vs. plunger gate: C$_\text{C}$/ C$_\text{P}\,=\,4$. (c) Left and (d) right perimeter-defining gates: C$_\text{L}$/ C$_\text{P}\,=\,3.8$ and  C$_\text{R}$/ C$_\text{P}\,=\,4.3$. }
	\label{fig2}
\end{figure}
\newpage
\begin{figure}[h!]
	\includegraphics[scale=1]{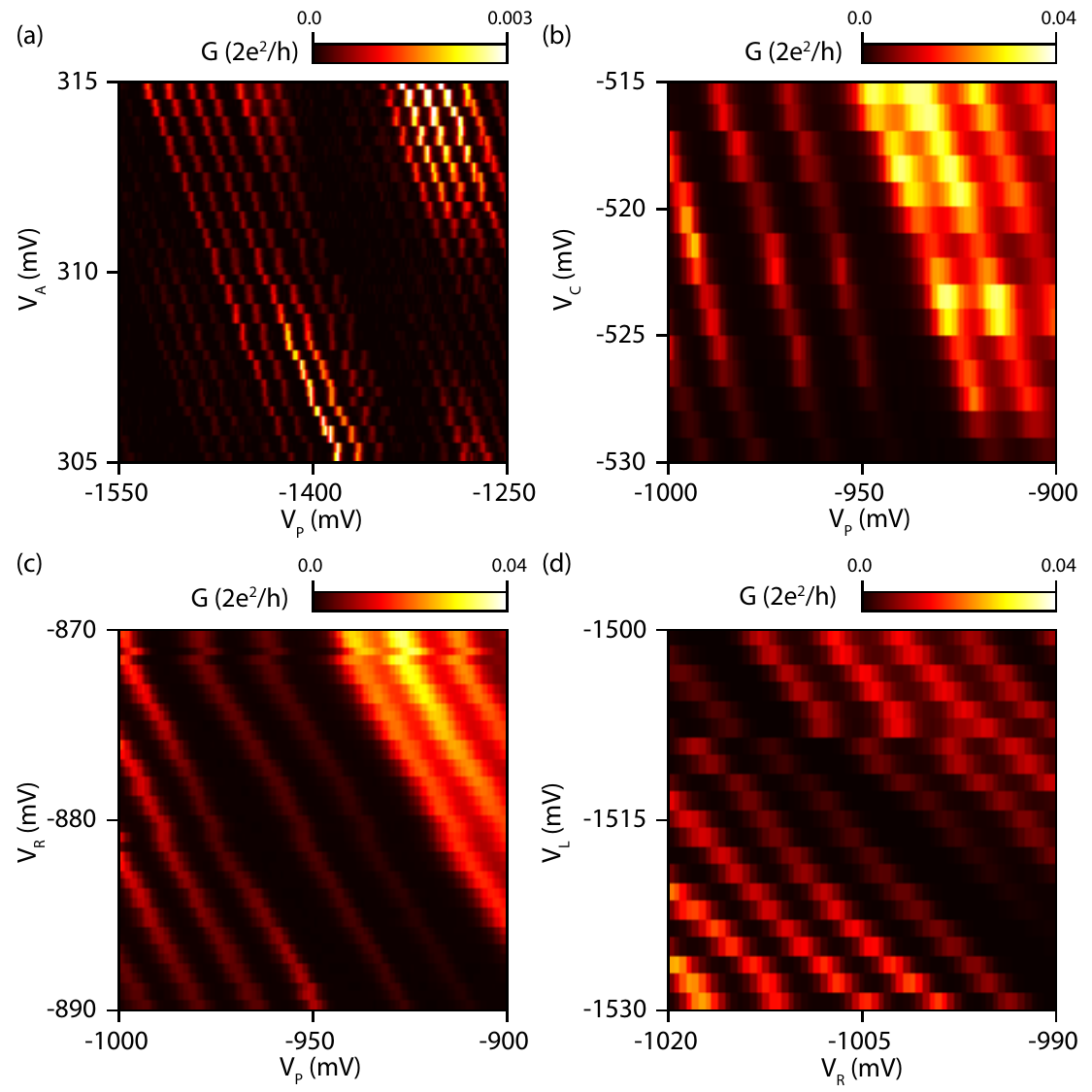}
	\caption{Conductance maps for S2 double-layer dot at zero bias as  a function of two specific gates. S2 has the same type of design as S1 but made on a different wafer. (a) Accumulation gate vs. plunger gate: C$_\text{A}$/ C$_\text{P}\,=\,11$. Compared to S1, the accumulation gate contribution to total capacitance is increased here, which can be partially attributed to the thinner quantum well barrier. (b) Central gate vs. plunger gate: C$_\text{C}$/ C$_\text{P}\,=\,2$. (c) Right gate vs. plunger gate: $\text{C}_\text{R} /\, \text{C}_\text{P}\,=\,2.6$. (d) Relative coupling of left and right gates: C$_\text{L}$/ C$_\text{R}\,=\,1$.}
	\label{fig3}
\end{figure}
\newpage
\begin{figure}[h!]
	\includegraphics[scale=1]{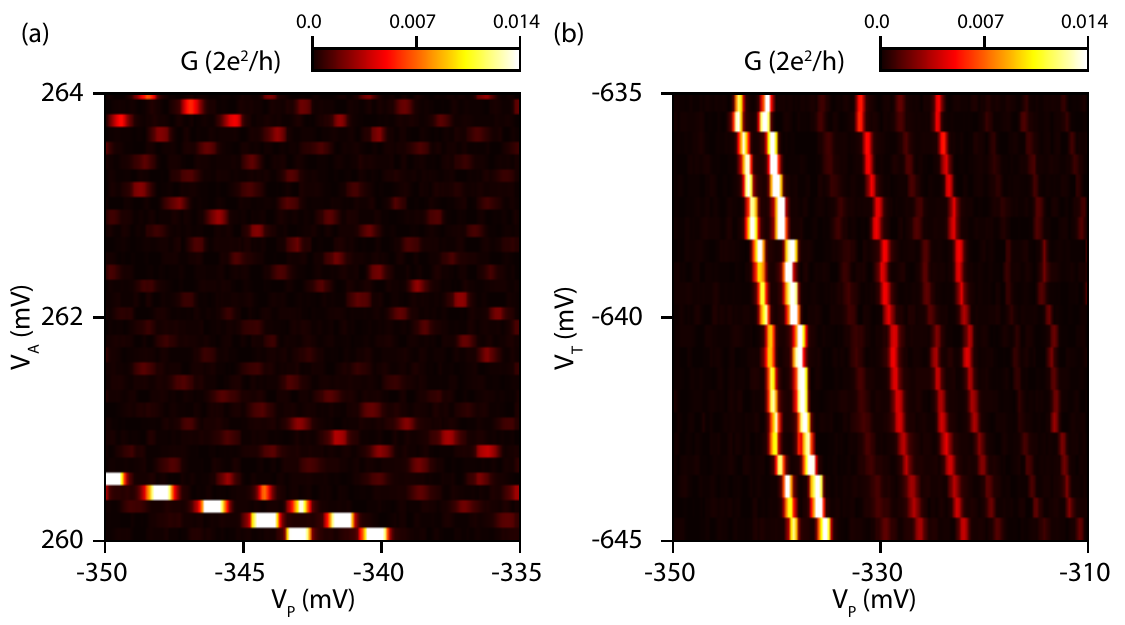}
	\caption{Conductance maps for S3 single-layer dot at zero bias as a function of two specific gates. (a) Accumulation gate vs. plunger gate: C$_\text{A}$/ C$_\text{P}\,=\,12$. (b) Tunnel gate T vs. plunger gate: C$_\text{T}$/~C$_\text{P}\,=\,0.6$.}
	\label{fig4}
\end{figure}
\vspace{2cm}
\begin{table}[h!]
	\centering
	\begin{tabular}{ |P{1.2cm}||P{1.4cm}|P{1.4cm}|P{1.4cm}|P{1.4cm}|P{1.4cm}|  }
		\hline
		Device &  C$_\text{A}$/C$_\text{P}$ &  C$_\text{C}$/C$_\text{P}$ &  C$_\text{T}$/C$_\text{P}$ &  C$_\text{L}$/C$_\text{P}$ &  C$_\text{R}$/C$_\text{P}$\\
		\hline
		S1	 & 5.5 & 4 & NA & 3.8 &  4.3\\
		S2	 & 11 & 2 & NA & 2.6  &   2.6\\
		S3	 & 12 & NA & 0.6 & NA & NA\\
		\hline
	\end{tabular}
	\caption{Capacitance ratios for the quantum dots S1, S2, S3. Parameters, irrelevant for a particular design, indicated as NA.}
	\label{tableCR}
\end{table}

\clearpage
\section{Effective radius estimations}

To estimate the effective radius, r$_\text{eff}$, we use a planar capacitor model for the contribution of the accumulation gate. Since there are two layers of different materials (with the thickness $\text{d}_1$ for the top barrier and  $\text{d}_2$ for the dielectric layer) between the confined system and the top gate, accumulation gate capacitance is determined as follows: $\text{C}_\text{A} = \epsilon_0 S_{\text{eff}} \left( \frac{\text{d}_1}{\epsilon_{\text{AlInSb}}} + \frac{\text{d}_2}{\epsilon_{\text{AlOx}}} \right)^{-1}$, where $S_{\text{eff}} = \pi \text{r}_{\text{eff}}^2$. The absolute capacitance value obtained from the charging energy, plunger lever arm and the ratio of accumulation and plunger gate capacitance: $\text{C}_\text{A} = \frac{e^2}{\text{E}_{\text{c}}} \cdot \frac{\text{C}_{\text{P}}}{\text{C}_{\text{tot}}} \cdot \frac{\text{C}_{\text{A}}}{\text{C}_{\text{P}}}$. Using measured parameters of the dots and $\epsilon_{\text{AlInSb}} = 17$ \cite{dixon1980measurement}, $\epsilon_{\text{AlOx}} = 9$ \cite{biercuk2003low}, we can estimate the effective radius, see Table~\ref{table1}. For double-layer dots S1 ans S2 effective radius is
lower than the lithographic dimensions, which can be attributed to the depletion
regions around the fine gates.

\begin{table}[h!]
	\centering
	\begin{tabular}{ |P{1.2cm}||P{0.85cm}|P{0.85cm}|P{1.0cm}|P{1.1cm}|P{0.85cm}|P{1.4cm}|P{1.4cm}|P{0.85cm}|P{1.0cm}|  }
		\hline
		Device & d$_1$ (nm) & d$_2$ (nm) & gate layers & E$_\text{c}$ (meV) &  C$_\text{tot}$ (aF) &  C$_\text{tot}$/C$_\text{P}$ & C$_\text{tot}$/C$_\text{A}$ & r$_\text{eff}$ (nm)&r$_\text{design}$ (nm)\\
		\hline
		S1	 & 40 & 80 & 2 & 1.23 &  130 & 29 & 5.5 &100& 130\\
		S2	 & 10 & 80 & 2 & 0.5    &   320 & 33 & 11 & 200 & 260\\
		S3	 & 10 & 40 & 1 & 0.16    & 1000 & 22.4& 11.6 & 310& 300\\
		\hline
	\end{tabular}
	\caption{The extracted parameters for the quantum dots S1, S2, S3.}
	\label{table1}
\end{table}

\clearpage
\section{Additional measurements for the device S1}

\begin{figure}[h!]
	\includegraphics[scale=1]{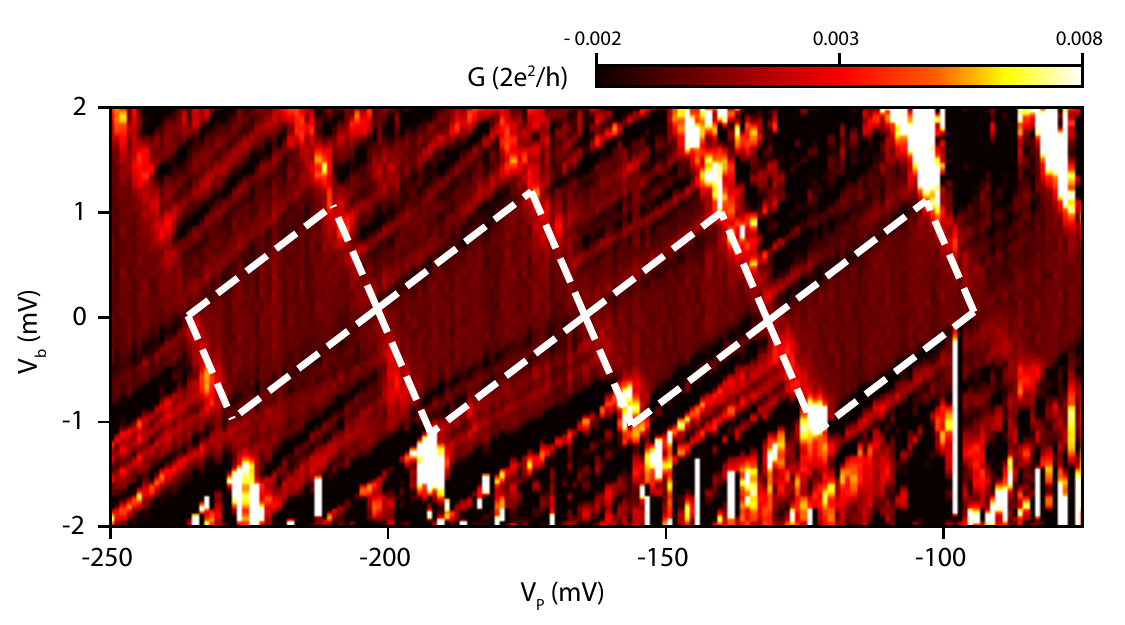}
	\caption{Coulomb diamonds for the S1, corresponding to the gate configuration of Fig.~4 in main text. Gates settings,  V$_\text{L} = -202$ mV, V$_\text{R} = -252$ mV, V$_\text{C} = -145$ mV, V$_\text{A} = 500$ mV, are the same as for the magnetic field scan in the Fig4a of the main text. Diamond parameters are used to extract the plunger lever arm C$_\text{tot}$/ C$_\text{P}\,=\,$ 33. Even-odd periodicity in the diamond size is apparent, resulting from the double degeneracy of the orbital levels.}
	\label{fig5}
\end{figure}

\clearpage
\bibliography{SI}